\documentclass{article}

\usepackage{pdfpages} 
\usepackage{fancyhdr} 

\usepackage{amsmath,amssymb,amsfonts,amsthm}
\usepackage{mathtools}
\usepackage{graphicx}
\usepackage{tikz}
\usepackage{float}
\usepackage{bm}
\usepackage{url}
\usepackage{underscore}
\usepackage{colortbl}
\usepackage{tabularx}
\usepackage{fancyhdr}
\usepackage[normalem]{ulem}
\usepackage{textcomp}
\usepackage{xcolor}
\usepackage{ascmac}
\usepackage{multicol}
\usepackage{wrapfig}
\usepackage{algorithmic}
\usepackage{appendix}
\usepackage{comment}
\usepackage[utf8]{inputenc}
\usepackage[T1]{fontenc}
\usepackage{booktabs}
\usepackage[left=30mm,right=30mm]{geometry}
\usepackage{booktabs}
\usepackage{enumitem}

\theoremstyle{definition}

\newtheorem{theorem}{Theorem}
\newtheorem{lemma}{Lemma}

\newcommand{\uir}[1]{\textsc{UIR$_{#1}$}}
\newcommand{\pa}{\textit{Player A}}
\newcommand{\pb}{\textit{Player B}}
\newcommand{\uvg}{\textsc{Undirected vertex geography}}
\newcommand{\dir}{\textsc{DIR}}
\newcommand{\classP}{P}

\fancypagestyle{centerednumber}{
  \fancyhf{}
  \fancyfoot[C]{\thepage}
  \renewcommand{\headrulewidth}{0pt}
}

\title{The computational complexity of Undirected Vertex Geography with maximum height $3$}
\author{Craig Tennenhouse, Kanae Yoshiwatari}
\date{}

\begin{document}

\maketitle

\abstract{
The impartial game \textsc{Undirected Vertex Geography} is known to be polynomial time solvable. The variant in which vertices are assigned heights with maximum value $m$ that reduce by $1$ whenever the token is moved from that vertex, denoted \uir{m}, is known to be polynomial time solvable when $m=2$ and PSPACE-complete when $m=4$. We solve the open question of the computational complexity of \uir{3} by a reduction from \textsc{Directed vertex geography}, and show that the complexity holds when the underlying graph is planar and (a) has degree at most $3$, (b) is bipartite with maximum degree $4$, or (c) is bipartite with maximum degree $5$ and only has heights of $1$ and $3$.
}

\section{Introduction}
The impartial game \textsc{Generalized Geography}, which we will denote by \dir, is played on a directed graph $D$. Players alternate moving a token from a vertex in $D$ to an out-neighbor, removing the former vertex in the process. Under normal play rules, the last player to make a legal move wins. The name comes from the traditional game that this ruleset models, where players take turns choosing the name of a city whose first letter is the same as the last letter of the last named city. If each vertex is a city name, then a directed edge represents a legal play from that city name to another. \dir\ is known to be PSPACE-complete by a reduction from \textsc{The Formula Game}, even when the underlying graph of $D$ is bipartite, planar, and has maximum degree $3$~\cite{lichtenstein1980go}. 

The undirected variant \uvg\ is played on a connected undirected graph $G$. \uvg\ was proven to be polynomial time solvable in 1993 by Fraenkel, Scheinerman, and Ullman~\cite{fraenkel1993undirected}. A number of variants of \dir\ and \uvg\ have been studied. One such variant, \uir{m}, is played on an undirected graph with positive integer heights of at most $m$ assigned to the vertices. In this variant, every move from a vertex lowers the height by $1$, vertices whose height reaches $0$ are removed, and no move may be made from a vertex of height $h$ to a neighbor with height strictly less than $(h-1)$. In~\cite{fox2022computational} it is shown that \uir{2}\ is in \classP\, and that \uir{4}\ is PSPACE-complete even when restricted to bipartite graphs with maximum degree $3$. 

In this manuscript we resolve the open question of the computational complexity of \uir{3}\ by direct reductions from \dir, and provide various degree and partite restrictions.

\section{Preliminaries}

For a graph $G$, we denote by $V(G)$ and $E(G)$ the vertex set and the edge set of $G$, respectively,
and by $|V(G)|$ and $|E(G)|$ the number of vertices and the number of edges of $G$, respectively.

\section{PSPACE-completeness}
\begin{figure}[ht]
\centering
\includegraphics[height=.4\textwidth]{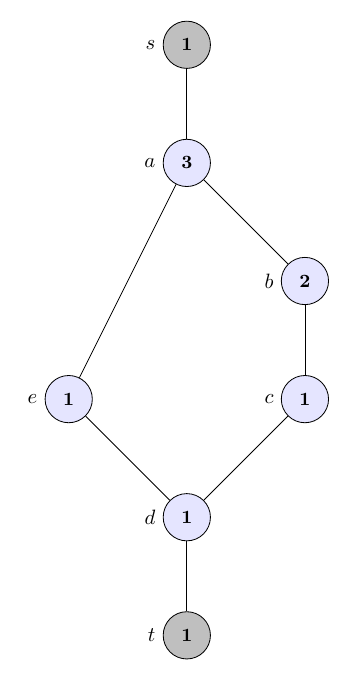}
\includegraphics[height=.4\textwidth]{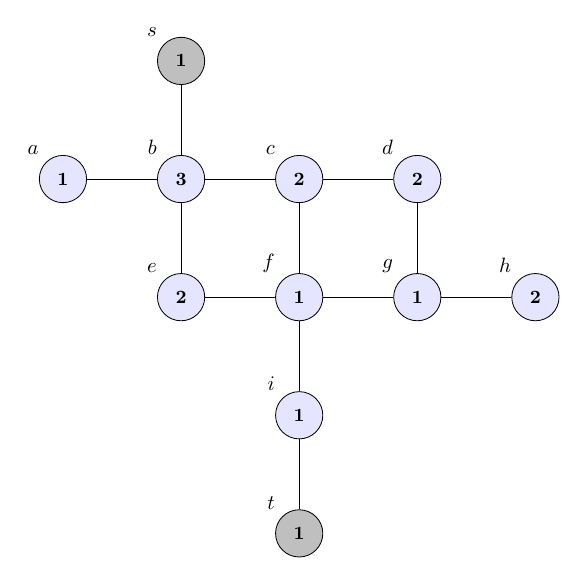}
\includegraphics[height=.4\textwidth]{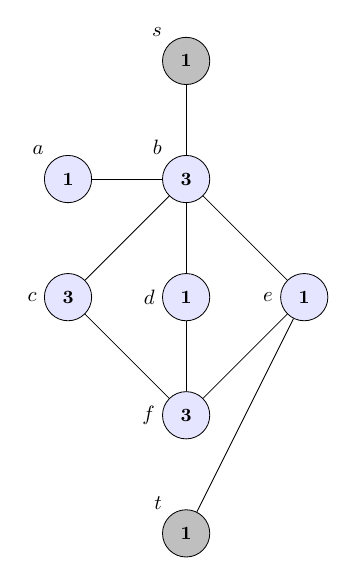}
\caption{The edge gadgets $T$, $B$, and $O$}
\label{fig:gadgets}
\end{figure}

Each of our reductions from \dir\ consists of a single edge gadget. That is, given an instance of \dir, say the directed graph $D$ with the token at vertex $v$, an undirected graph $G(D,v)$ can be constructed in which $V(D) \subsetneq V(G)$ (or $V(D)=V(G)$ in the trivial case with no edges), the position of the token remains the same, and for every directed edge $st$ in $E(D)$ there is a connected undirected graph with vertex heights in $G$. In Figure~\ref{fig:gadgets} we show the three edge gadgets $T, B, O$ that we consider in this manuscript.

The edge gadgets $T, B, O$ in Figure~\ref{fig:gadgets} are used for reductions from \dir\ to maintain maximum degree $3$, bipartiteness with maximum degree four, and bipartiteness  with only weights $1$ and $3$, respectively. We prove here that each gadget can be used to replace each directed edge $st$ in a game $(D,v)$ of \dir, and that the resulting game of \uir{3}\ is a next player win if and only if $(D,v)$ is also a next player win.

\begin{lemma}\label{lem:gadgets}
The next player to move on gadget $T$, $B$, or $O$ from vertex $s$ can reach vertex $t$. Moreover, both players can prevent the other from winning inside the gadget. Finally, if the token is instead on vertex $t$ at the start, then the next player to move loses in the gadget. We refer to this final property as the \emph{diode property}.  
\end{lemma}

Note that we require the diode property to ensure that no player can move from $u$ to $v$ in \uir{3}\ if $vu$ is a directed edge in $D$ but $uv$ is not.
The proof of Lemma~\ref{lem:gadgets} requires difficult to track case analyses. Therefore, we have chosen to include the proof in the form of game trees in the appendix to this manuscript. 

Now we move on to the main result of this manuscript.

\begin{theorem}\label{thm:main}
The game \uir{3}\ is PSPACE-complete, even when restricted to
\begin{itemize}
    \item[(a)] Planar graphs with degree at most $3$
    \item[(b)] Planar bipartite graphs with degree at most $4$
    \item[(c)] Planar bipartite graphs with degree at most $5$ and only heights $1$ and $3$
\end{itemize}
\end{theorem}

\begin{proof}
Consider the \dir\ game $(D,v)$ where $D$ has maximum degree $3$ and is planar and bipartite, as in~\cite{lichtenstein1980go}. Let $(G,v)$ be an instance of \uir{3}\ where we construct $G$ by including $V(D)$, and then replace every directed edge $st$ in $D$ with the gadget $T, B,$ or $O$. By Lemma~\ref{lem:gadgets}, under optimal play by both players the token at vertex $v$ will reach an out-neighbor of $v$ in an odd number of turns. Any attempt by the next player to move backwards, from $v$ to an in-neighbor, results in a previous player win. Therefore, a move in $(D,v)$ is equivalent to a move in $(G,v)$, and $(D,v)$ is a next player winning position in \dir\ if and only if $(G,v)$ is a next player winning position in \uir{3}. The construction of $(G,v)$ takes polynomial time in $E(D)$. Hence, determining whether or not a position in \uir{3}\ is a next player win is PSPACE-hard. 

The number of turns in $(G,v)$ is at most $|E(D)|k + |V(D)|$,
where $k$ is the sum of heights in the gadget used, excluding the heights of $s$ and $t$. 
Therefore, \uir{3}\ is in PSPACE, and thus the game is PSPACE-complete.
\end{proof}

\section{Acknowledgments}

The authors wish to thank MEXT Supporting Pioneering Research through AI for 1,000 Discovery challenges Program (SPReAD) Japan Grant Number JPMXP1726301948 for their generous support of this research.


\section{Appendix}

Below are the game trees in place of a written proof of Lemma~\ref{lem:gadgets}. We denote by \pa\ the next player to move, and by \pb\ the previous player. For each gadget we show the following three things are true.
\begin{itemize}
    \item[(i)] If \pa\ begins at vertex $s$ then \pa\ can avoid losing in the gadget.
    \item[(ii)] If \pa\ begins at vertex $s$ then \pb\ can avoid losing on a non-$t$ node.
    \item[(iii)] If \pa\ begins at vertex $t$ then \pb\ will win in the gadget. 
\end{itemize}

On the following pages, each tree is titled to indicate which gadget and which of the three above properties it addresses. Branching is only shown to address all possible opponent choices, and we only include optimal moves for the relevant player to meet each requirement.

\clearpage
\pagestyle{fancy}
\fancyhf{}
\fancyfoot[R]{\thepage}
\renewcommand{\headrulewidth}{0pt}


\includepdf[
    pages={1},
    pagecommand={
        \vspace*{-1.5cm}
        \noindent\emph{\pa\ can avoid losing in $T$}
        \vspace{0.5cm}
    },
    scale=.95,      
    offset=0 -1.8cm
]{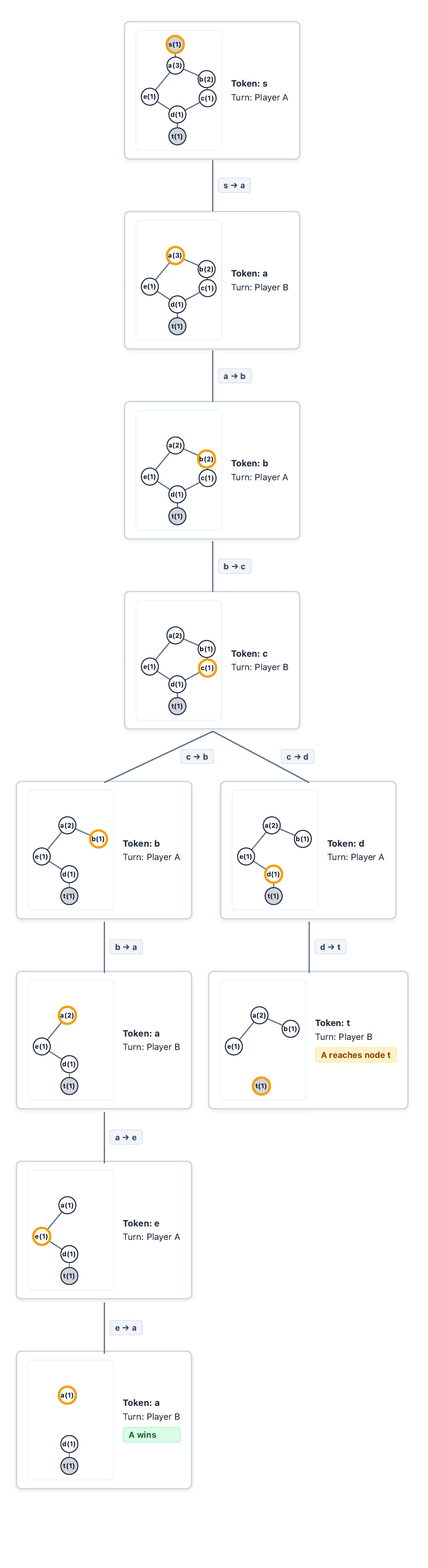}

\includepdf[
    pages={1},
    pagecommand={
        \vspace*{-1.5cm}
        \noindent\emph{\pb\ can avoid losing in $T\setminus \{t\}$}
        \vspace{0.5cm}
    },
    scale=.95,      
    offset=0 -1.8cm
]{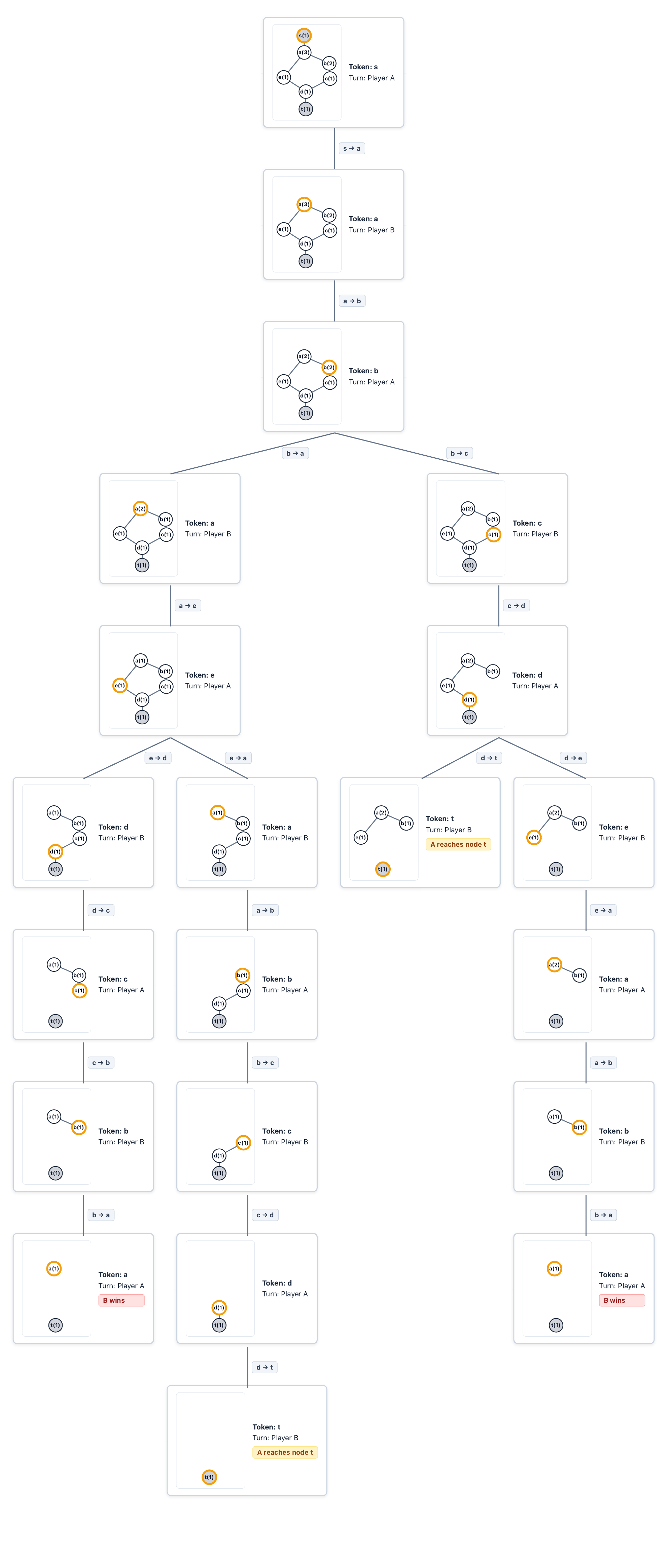}

\includepdf[
    pages={1},
    pagecommand={
        \vspace*{-1.5cm}
        \noindent\emph{Diode property for gadget $T$}
        \vspace{0.5cm}
    },
    scale=.95,      
    offset=0 -1.8cm
]{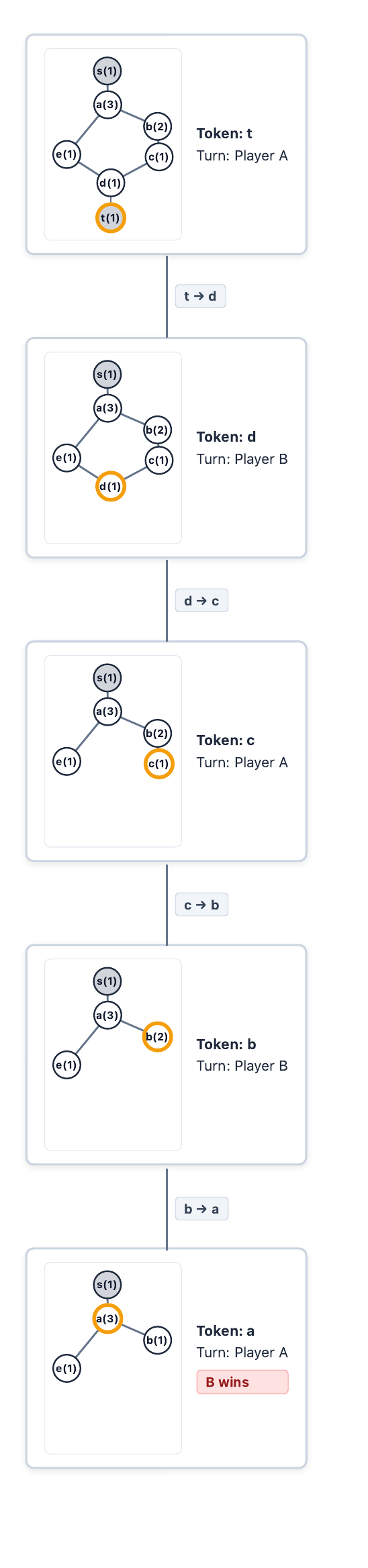}


\includepdf[
    pages={1},
    pagecommand={
        \vspace*{-1.5cm}
        \noindent\emph{\pa\ can avoid losing in $B$}
        \vspace{0.5cm}
    },
    scale=.95,      
    offset=0 -1.8cm
]{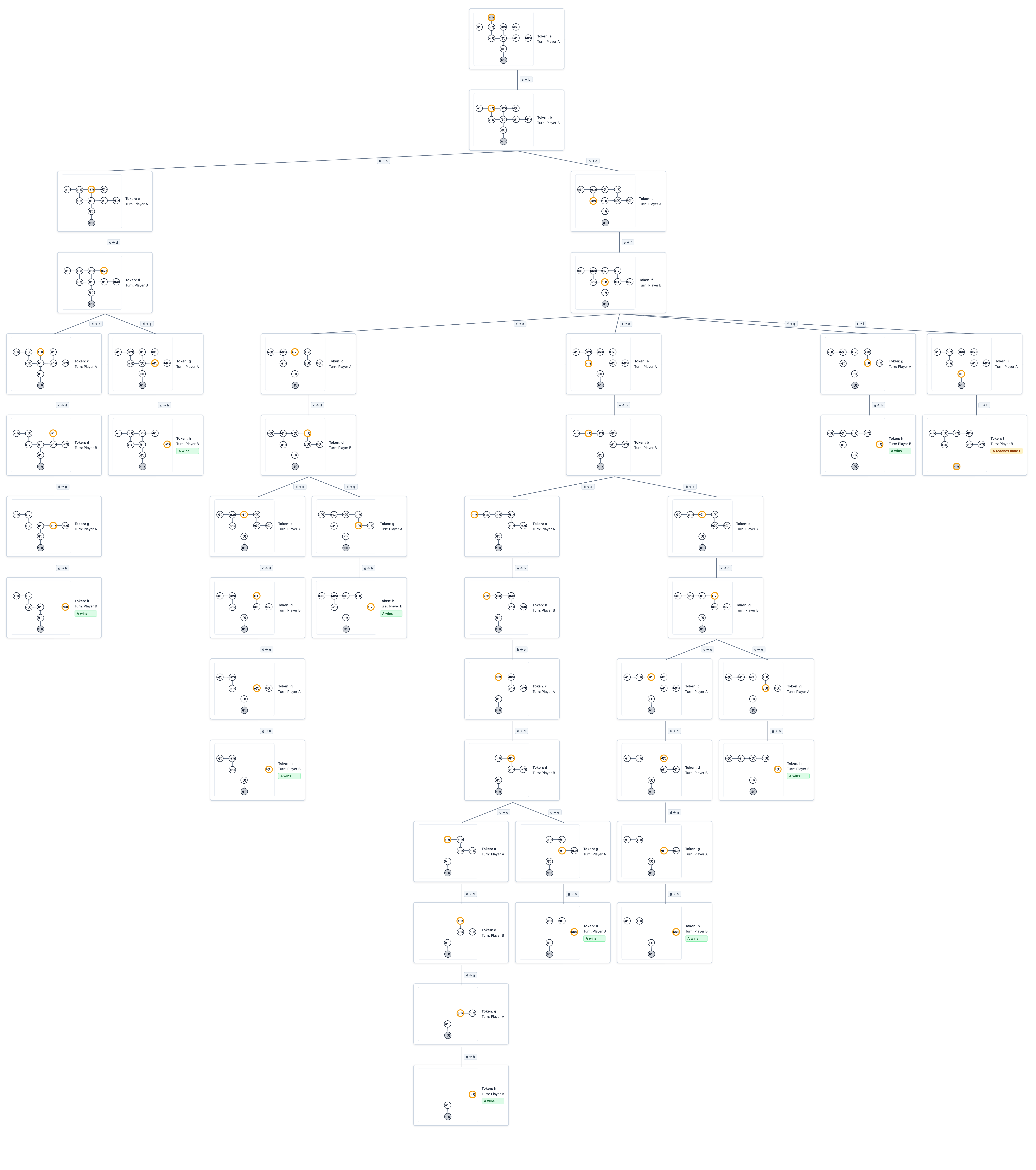}

\includepdf[
    pages={1},
    pagecommand={
        \vspace*{-1.5cm}
        \noindent\emph{\pb\ can avoid losing in $B\setminus \{t\}$}
        \vspace{0.5cm}
    },
    scale=.90,      
    offset=0 -1.8cm
]{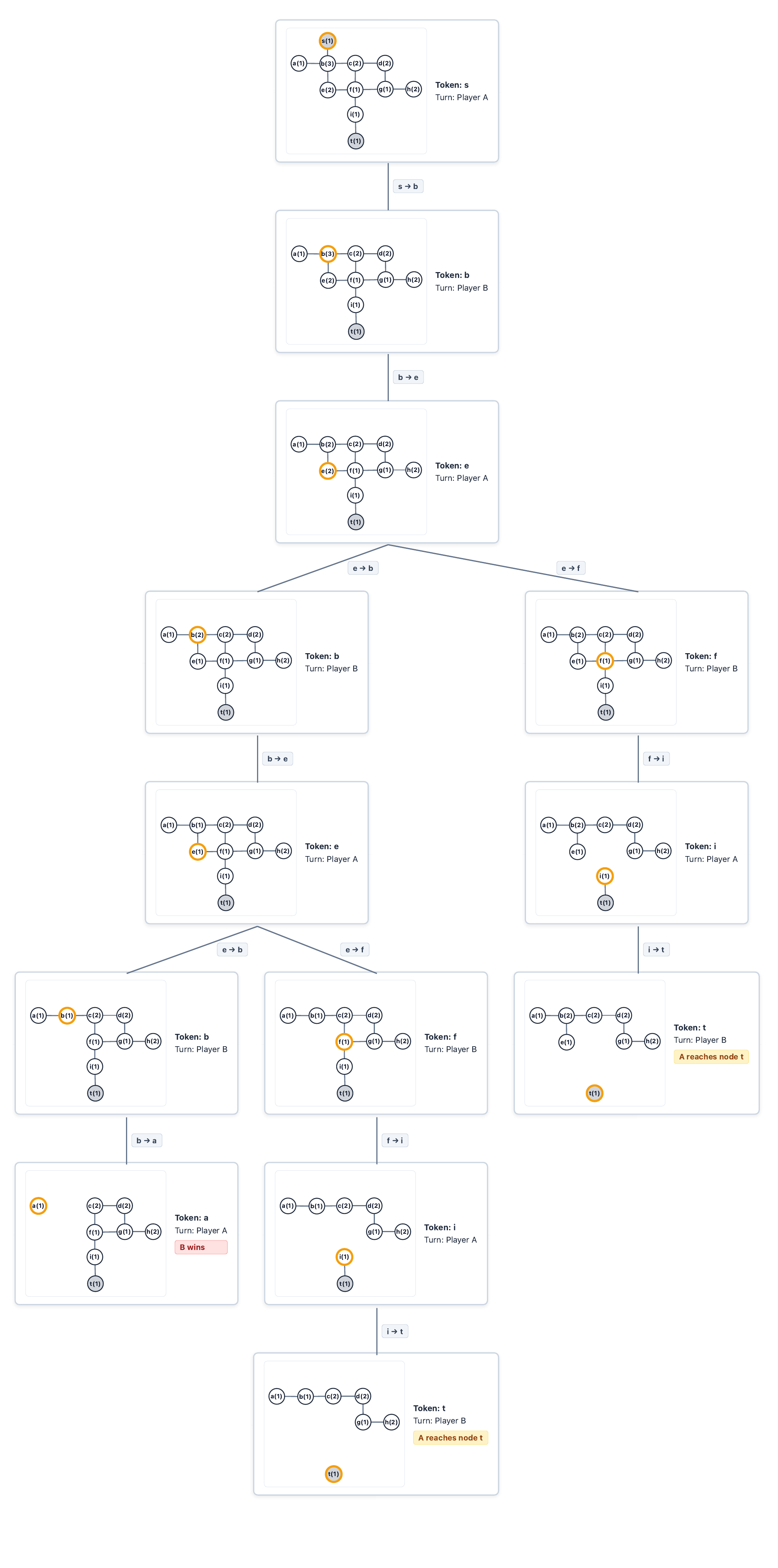}

\includepdf[
    pages={1},
    pagecommand={
        \vspace*{-1.5cm}
        \noindent\emph{Diode property for gadget $B$}
        \vspace{0.5cm}
    },
    scale=.90,      
    offset=0 -1.8cm
]{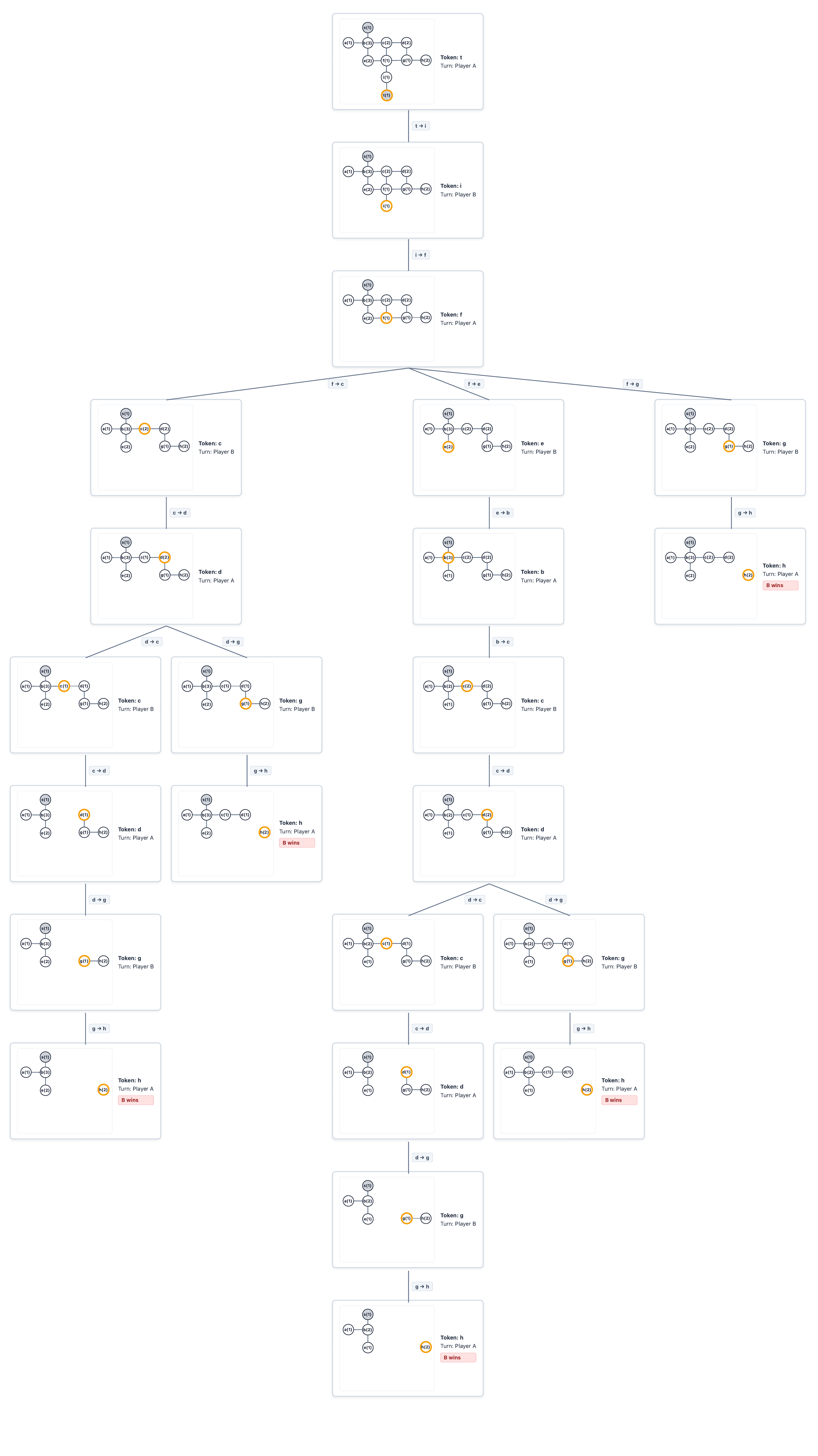}


\includepdf[
    pages={1},
    pagecommand={
        \vspace*{-1.5cm}
        \noindent\emph{\pa\ can avoid losing in $O$}
        \vspace{0.5cm}
    },
    scale=.95,      
    offset=0 -1.8cm
]{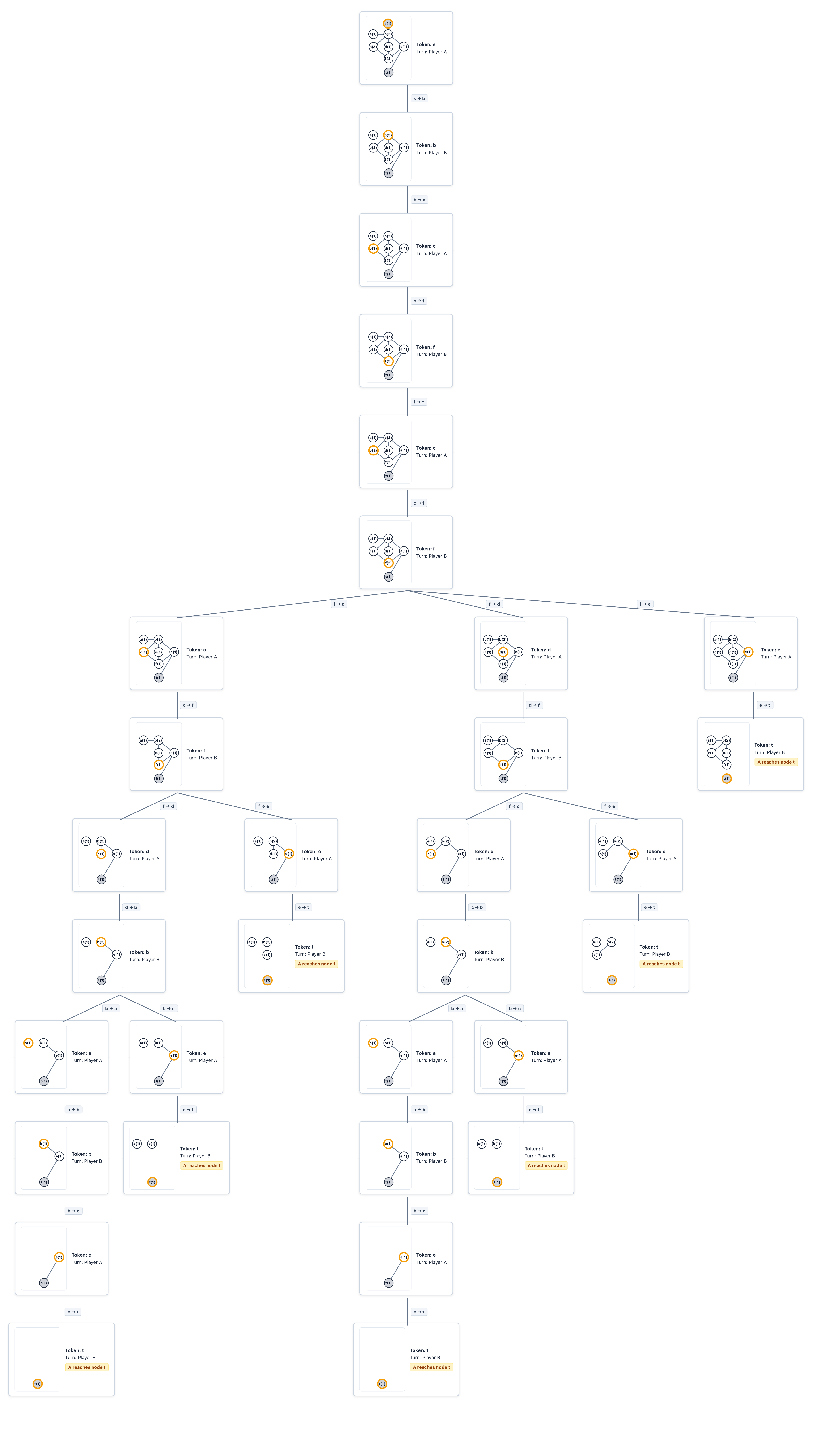}

\thispagestyle{centerednumber}

\includepdf[
    pages={1},
    pagecommand={
        \vspace*{-1.5cm}
        \noindent\emph{\pb\ can avoid losing in $O\setminus \{t\}$}
        \vspace{0.5cm}
    },
    scale=.95,      
    offset=0 -1.8cm
]{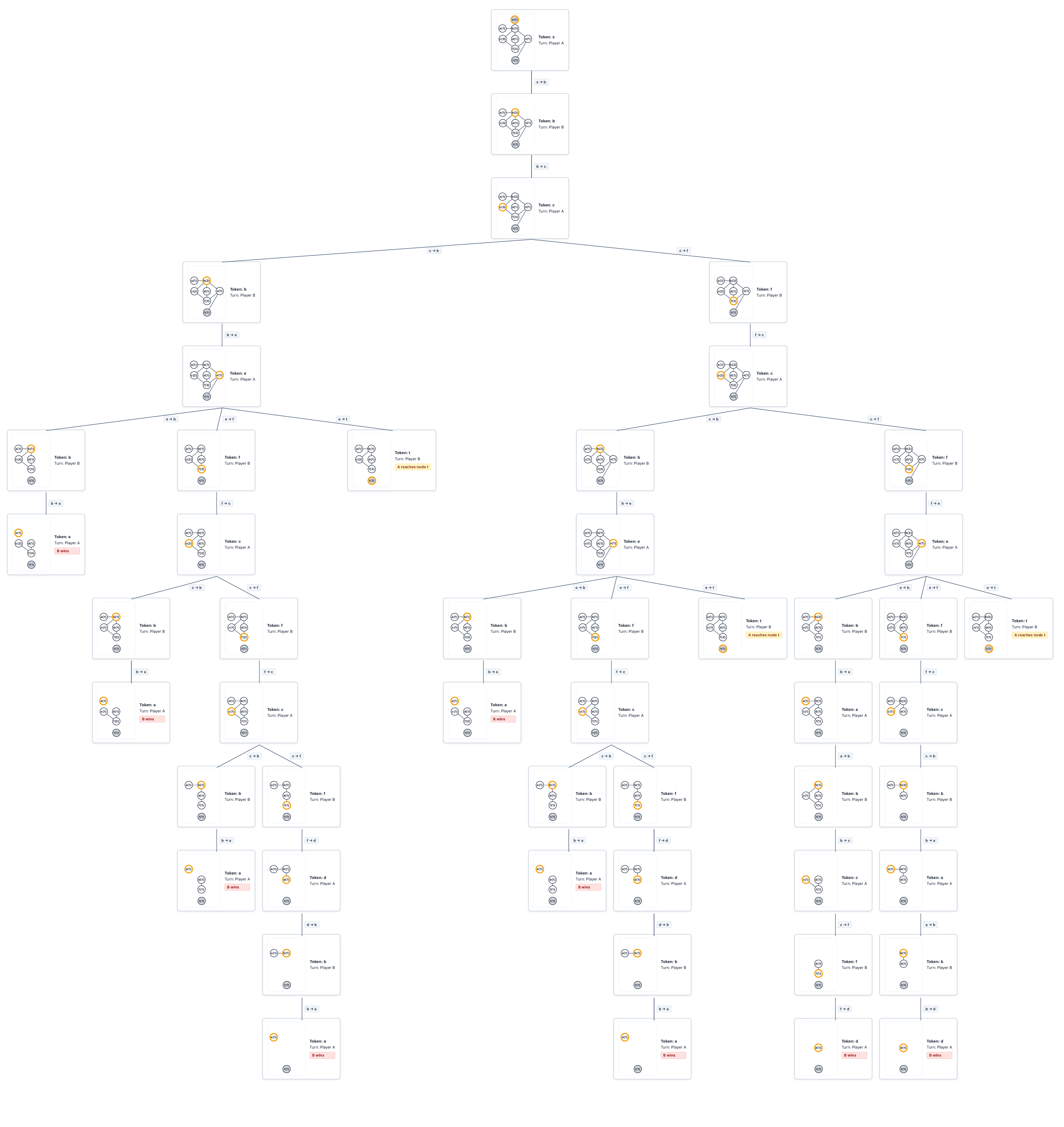}

\includepdf[
    pages={1},
    pagecommand={
        \vspace*{-1.5cm}
        \noindent\emph{Diode property for gadget $O$}
        \vspace{0.5cm}
    },
    scale=.90,      
    offset=0 -1.8cm
]{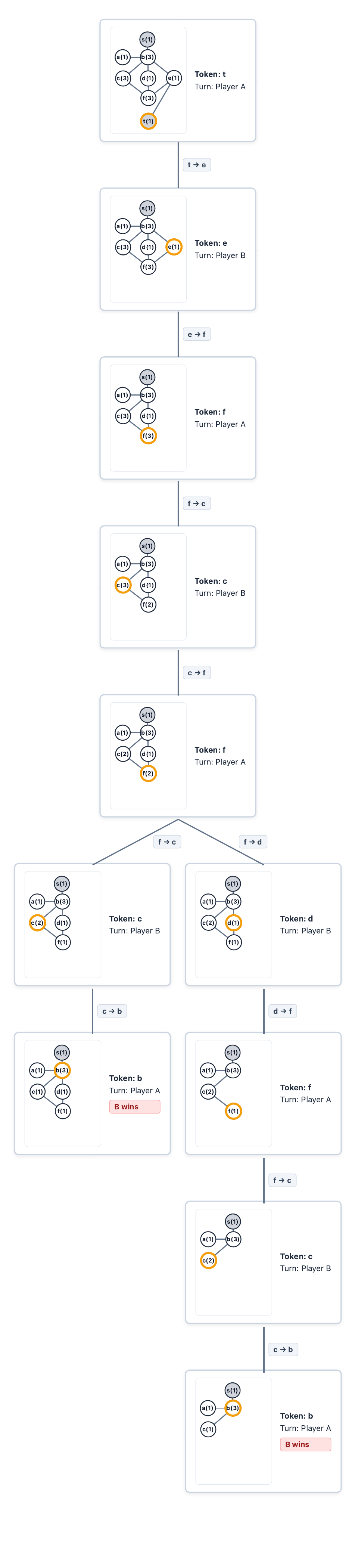}

\thispagestyle{centerednumber}

\bibliographystyle{plain}
\bibliography{sankou}

\end{document}